\documentclass[Afour,sageh,times]{sagej}

\usepackage{moreverb,url}
\usepackage{siunitx}

\usepackage[colorlinks,bookmarksopen,bookmarksnumbered,citecolor=red,urlcolor=red]{hyperref}

\usepackage{xcolor}

\newcommand\BibTeX{{\rmfamily B\kern-.05em \textsc{i\kern-.025em b}\kern-.08em
T\kern-.1667em\lower.7ex\hbox{E}\kern-.125emX}}

\def\volumeyear{2016}

\newcommand\hil[1]{%
  \bgroup
  \hskip0pt\color{red!80!black}%
  #1%
  \egroup
}

\newif\ifhighlight
\highlightfalse
\ifhighlight
    
\else
    \renewcommand{\hil}[1]{#1}
\fi

\begin{document}

\runninghead{Sawant et al}

\title{Classification of Various Types of Damages in Honeycomb Composite Sandwich Structures using Guided Wave Structural Health Monitoring
}

\author{Shruti Sawant\affilnum{1}, Jeslin Thalapil\affilnum{1}, Siddharth Tallur\affilnum{1}, Sauvik Banerjee\affilnum{2} and Amit Sethi\affilnum{1}}

\affiliation{\affilnum{1}Department of Electrical Engineering,Indian Institute of Technology Bombay,Mumbai,400076,Maharashtra,India\\
\affilnum{2}Department of Civil Engineering,Indian Institute of Technology Bombay,Mumbai,400076,Maharashtra,India}

\corrauth{Amit Sethi,
Department of Electrical Engineering,
Indian Institute of Technology Bombay,
Mumbai,
400076,
Maharashtra,
India.}

\email{asethi@iitb.ac.in}

\begin{abstract}
Classification of damages in honeycomb composite sandwich structure (HCSS) is important to decide remedial actions. However, previous studies have only detected damages using deviations of monitoring signal from healthy (baseline) using a guided wave (GW) based structural health monitoring system. Classification between various types of damages has not been reported for challenging cases. We show that using careful feature engineering and machine learning it is possible to classify between various types of damages such as core crush (CC), high density core (HDC), lost film adhesive (LFA) and teflon release film (TRF). \hil{We believe that we are the first to report numerical models for four types of damages in HCSS, which is followed up with experimental validation.} We found that two out of four damages affect the GW signal in a particularly similar manner. We extracted and evaluated multiple features from time as well as frequency domains, and also experimented with features relative to as baseline as well as those that were baseline-free. Using Pearson's correlation coefficient based filtering, redundant features were eliminated. Finally, using an optimal feature set determined using feature elimination, high accuracy was achieved with a random forest classifier on held-out signals. \hil{For evaluating performance of the proposed method for different damage sizes, we used simulated data obtained from extensive parametric studies and got an accuracy of 77.89\%.}  Interpretability studies to determine importance of various features showed that features computed using the baseline signal prove more effective as compared to baseline-free features. 

\end{abstract}


\maketitle

\section{Introduction}
Honeycomb composite sandwich structure (HCSS) consists of a lightweight honeycomb core sandwiched between two composite skins. HCSS is utilized in a variety of applications, including aerospace, because of their high strength-to-weight ratio, high energy absorption, and outstanding thermal and acoustic insulation. However, during manufacturing or operations HCSS can get damaged in various ways which can jeopardize structural integrity \cite{shipsha2005compression,jollivet2013damage,balasko2005classification,SIKDAR2019107195}. Damage classification is important because classifying the type of damage occurring within a structure can allow the end user to assess what kind of repairs, if any, that a component requires. 

Ultrasonic guided wave-based structural health monitoring systems (GW-SHMs) are popular for a variety of composite structures due to low cost of instrumentation, sensitivity to small defects, and ability of guided waves (GWs) to travel long distances on the structure without attenuation \cite{yan2010ultrasonic,abbas2018structural}. Various types of damages affect GW signals in multiple ways. Disbond in HCSS causes increase in amplitude of A0 mode, while high density core (HDC) results in a decrease in amplitude \cite{sikdar2016identification}. However, due to a large variety of damages that may occur throughout life-cycle of the material, different types of damages may result in similar changes in amplitude and phase of GW signals. Perhaps due to this challenge, damage classification has not been explored well enough. 

Although damage classification in structures using GW-SHM has been attempted before, these studies have various limitations with respect to the goals of the current study, which is to classify damages in HCSS. For instance, previously we reported high accuracy for damage classification for three types of damages -- rivet hole, notch, and added mass -- using convolutional neural networks (CNNs) under various temperatures but in a homogeneous aluminum panel \cite{sawant2022temperature}. We used simulated data for the three damages along with experimental validation of binary classification for the notch damage. 
Tibaduiza et al proposed hierarchical nonlinear principal component (PCA) analysis approach for classifying delamination and cracking of the skin in  a carbon fibre reinforced polymer (CFRP) sandwich structure \cite{tibaduiza2018damage}.

Lee et al. \cite{LEE2022108148} demonstrated use of unsupervised deep auto encoder (DAE). They use two different thresholds on reconstruction error computed at output of DAE to distinguish between matrix cracking and delamination however there is no report of performance parameters, size or architecture of DAE. Deep learning-based methods reported for GW-SHM of composite structure in supervised \cite{RAUTELA2021106451} as well as unsupervised setups \cite{rai2022transfer} using public-domain Open Guided Waves (OGW) dataset \cite{moll2019open}. However, these studies model different locations of the added mass damage on CFRP plate as different classes for the classifier model. Therefore, these approaches actually solve damage localization problem for the same damage type, instead of classifying different damage types in composite structures.
In addition to trying damage classification, we wanted to avoid using deep learning based approaches in order to rein in the number of trainable parameters towards implementation on edge devices such as low-cost microcontrollers, field programmable gate arrays (FPGAs), and low-cost graphics processing units (GPUs).

GW-based damage characterization requires interpreting damage features that quantify deviation of monitoring signal from baseline signal. Conventional algorithms first extract the wavemode of interest from time-series data using group velocity, which requires knowledge of material properties, such as Young's modulus and density. Various statistical relations, such as sum of differences in each sample for wave mode of interest, difference in the peak amplitude of monitoring and baseline signal \cite{SAWANT2021106439}, correlation coefficient deviation (CCD) \cite{zhao2007active}, are used as features. Identifying effective damage features is crucial to the success of damage characterization. Selection of suitable features usually requires knowledge of material properties and varies for different structures and types of damages. The utility of easy-to-compute statistical metrics obtained from time-domain signals for machine fault diagnosis was shown by Bandyopadhyay et al. \cite{bandyopadhyay2018performance}. Elhariri et al. proposed hybrid filter-wrapper with multi-objective optimization feature selection method for crack severity recognition for machine health monitoring \cite{9083977}. A low-cost corrosion sensing system based on pulsed eddy current detection to detect the corrosion in steel reinforcing bars optimal feature selection and dimensionality reduction was demonstrated using principal component analysis \cite{9509426}. For GW-SHM system, feature engineering (FE) is still in nascent stages. Recently Liu et al. reported a feature selection method based on binary particle swarm optimization with a new fitness function proposed in combination with least-squares support-vector machine for damage identification in variety of challenging practical scenarios for switch rail damage \cite{liu2021multi}. FE approaches for GW-SHM of composite structures, especially for in-situ monitoring with ligher ML model suitable for edge implementation, remains to be explored.

In this work, we present a novel damage classification method using feature engineering for classification of four different types of damages in HCSS -- core crush (CC), Teflon release film (TRF), lack of film adhesive (LFA), and HDC. Two out of these damages (CC and LFA) cause increase in the amplitude over the baseline signal, whereas the other two result in the opposite effect. Choice and selection of the features was therefore important for this work. \hil{Firstly, we conducted numerical study for four damages using models developed in COMSOL. Secondly, we also conducted extensive parametric studies for different damage sizes for each damage type. It is costly and, in some cases, infeasible to generate experimental data for various sizes of damages. Therefore, to evaluate performance for such test cases, we carried out numerical simulations. Next, we collected data on HCSS using a pair of detachable contact-type Nano-30 transducers which was observed to be in accordance with numerical data. \emph{To the best of our knowledge, this is the first study to report numerical and experimental study for four different damage types in HCSS.}} Following noise augmentation of these experimental and simulated dataset, we computed ten statistical features that do not require knowledge of material properties, such as group velocity of various wavemodes or dispersion curve of the material, and compared baseline-free features as well as features computed using baseline in time and frequency domains. Using Pearson's correlation coefficient to filter redundant features we selected the optimal subset representing the most relevant information. Finally, we evaluated performance of five different classifiers using the optimal feature subset and observed that random forest classifier shows perfect accuracy using seven features on unseen experimental data. \hil{In the case of simulated data for different damage sizes, we achieved an accuracy of 77.89\% using random forest.} Our damage classification results are more accurate than those obtained using baseline-free features reported by Liu et al. \cite{liu2021multi}.

\begin{figure}[!ht]
\centering
\includegraphics[width=0.85\linewidth]{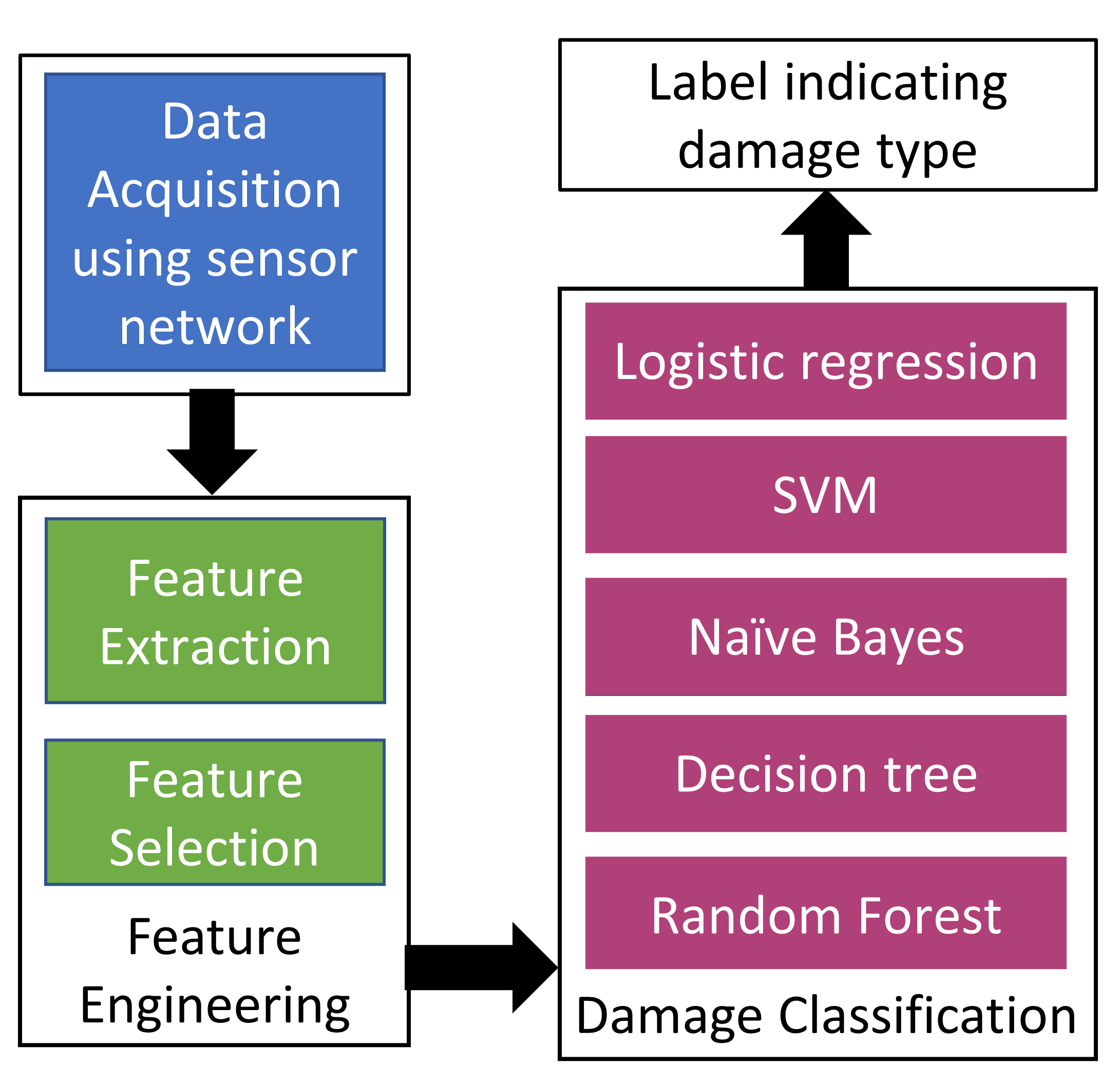}
\caption{Illustration of the proposed method using feature engineering for damage classification in GW-SHM systems.}
\label{fig:algo}
\end{figure}

\section{Feature engineering based damage classification method}\label{}
The proposed method mainly consists of four steps as illustrated in Figure \ref{fig:algo}:

\begin{itemize}
    \item Data acquisition: GW-SHM system for HCSS panel considered in this study uses a pair of detachable Nano-30 transducers to collect data non-invasively for three sensor paths around different types of damages on the panel. Preprocessing is essential as the experimental data contains offset.

    \item Feature extraction: Features play crucial role in quantifying deviation of monitoring data from the baseline data.  We computed a total of $10$ features in time as well as frequency domain for subsequent analysis. Along with these, we also explored baseline-free features in this study.
    
    \item Feature engineering: Primary feature identification involves selecting a subset of features to enhance the accuracy of the models and reduce generalization error introduced due to noise by irrelevant features. We used Pearson's correlation coefficient based filtering in this study using to select optimal set of features.
    
    \item Damage classification: Finally, to distinguish between four types of damage in HCSS panel, we trained classifier models. Using one-hot encoding scheme labels were generated for five classes (four damage types and baseline signal). Performance of five different classifier models was evaluated using the optimal feature subset. 
\end{itemize}

\begin{table*}[!tbp]
\begin{center}
\centering
\small
\caption{Features computed from time-domain GW signals. Glossary: $\bar{f}$: mean value of samples in time-series $f$, $T$: length of time-series, $f_b$: baseline signal, $A$: peak-to-peak amplitude of signal $f$, $A_b$: peak to peak amplitude of baseline signal $f_b$}

\label{tab:time_features}
\begin{tabular}{p{7cm}p{3.8cm}}
\hline
\textbf{Feature}      & \textbf{Expression} \\ \hline
\\ [0.5ex]

Correlation coefficient deviation (CCD) & ${1 - \sqrt{\frac{\{\int^{T}f_b(t)f(t)dt\}^2}{\{\int^{T}{f_b(t)}^2dt\int^{T}{f(t)}^2dt\}}}}$ \\ [3ex]

Mean absolute deviation (MAD) & $\frac{1}{T}\sum_{i=1}^{i=T}|f(i) - \bar{f}|$ \\ [3ex]

Normalised difference of signal energy (NSED)  & ${\frac{\int^T{|f(t)|^2}dt-\int^T{|f_{b}(t)|^2}dt}{\int^T{|f_{b}(t)|^2}dt}}$ \\ [3ex]

Peak to peak amplitude deviation (PPAD)  & ${A - A_{b}}$\\[1ex]

Root mean square (RMS) & $\sqrt{(1/T){\int^T {[f(t)]^2}dt}}$ \\ [3ex]

Root mean square deviation (RMSD) & $\sqrt{\frac{\int^T{[f(t)-f_{b}(t)]^2}dt}{\int^T{[f_{b}(t)]^2}dt}}$ \\ [3ex]

Spectral difference deviation (SDD) & $\sqrt{\frac{\{\int^{F}|F_b(f)-F(f)df|\}^2}{\int^{F}{F_b(f)}^2dt\int^{F}{F(f)}^2dt}}$\\ [3ex]

Ratio of signal energy (SER) & ${\frac{\int^T{|f(t)|^2}dt}{\int^T{|f_{b}(t)|^2}dt}}$ \\ [3ex]

Standard deviation ($\sigma$) & $\sqrt{\frac{\sum_{i=1}^{i=T}{\left(f(i)- \bar{f}\right)^2}}{T}}$ \\ [3ex]

Variance (VAR) & $\frac{\sum_{i=1}^{i=T}{\left(f(i) - \bar{f}\right)^2}}{T}$ \\ [3ex]


\hline
\end{tabular}
\end{center}
\end{table*}

\begin{table*}[!tbp]
\centering
\small
\caption{Material properties (standard notations) of CFRP lamina used in facesheet, adhesive, and honeycomb core in HCSS}
\label{tbl:scsprop} 
\resizebox{\textwidth}{!}{%
\begin{tabular}{ccccccccccc} 
\hline
\centering Material &
\multicolumn{1}{p{0.8cm}}{\centering $E_{11}$ (\SI{}{GPa})} &
\multicolumn{1}{p{0.8cm}}{\centering $E_{22}$ (\SI{}{GPa})} &
\multicolumn{1}{p{0.8cm}}{\centering $E_{33}$ (\SI{}{GPa})} &
\multicolumn{1}{p{0.8cm}}{\centering $G_{12}$ (\SI{}{GPa})} &
\multicolumn{1}{p{0.8cm}}{\centering $G_{23}$ (\SI{}{GPa})} &
\multicolumn{1}{p{0.8cm}}{\centering $G_{13}$ (\SI{}{GPa})} &
\centering $\nu_{12}$ &
\centering $\nu_{13}$ &
\centering $\nu_{23}$ &
\multicolumn{1}{p{1cm}|}{\centering $\rho$ (\SI{}{\kilo\gram \per \meter^3})} \\[0.5ex]
\hline 
\centering CFRP lamina  & $135$ & 9 & 9 & 4.5 & 3.103 & 4.5 & 0.29 & 0.29 & 0.45 & 1780\\[1ex]
\centering Adhesive  & 4.35 & 4.35 & 4.35 & 1.599 & 1.599 & 1.599 & 0.36 & 0.36 & 0.36 & 1100\\[1ex]
\centering Honeycomb core  & 0.028 & 0.028 & 0.40 & 0.074 & 0.103 & 0.221 & 0.318 & \SI{3.18e-5}{} & \SI{3.18e-5}{} & 36.8\\[1ex]
\hline
\end{tabular}
}
\end{table*}

\begin{table*}[!tbp]
\centering
\small
\caption{Material properties (standard notations) of damages (CC, HDC and TRF) considered for 2D numerical simulation of HCSS}
\label{tbl:damageprop} 
\resizebox{\textwidth}{!}{%
\begin{tabular}{ccccccccccc} 
\hline
\centering Material &
\multicolumn{1}{p{0.8cm}}{\centering $E_{11}$ (\SI{}{GPa})} &
\multicolumn{1}{p{0.8cm}}{\centering $E_{22}$ (\SI{}{GPa})} &
\multicolumn{1}{p{0.8cm}}{\centering $E_{33}$ (\SI{}{GPa})} &
\multicolumn{1}{p{0.8cm}}{\centering $G_{12}$ (\SI{}{GPa})} &
\multicolumn{1}{p{0.8cm}}{\centering $G_{23}$ (\SI{}{GPa})} &
\multicolumn{1}{p{0.8cm}}{\centering $G_{13}$ (\SI{}{GPa})} &
\centering $\nu_{12}$ &
\centering $\nu_{13}$ &
\centering $\nu_{23}$ &
\multicolumn{1}{p{1cm}|}{\centering $\rho$ (\SI{}{\kilo\gram \per \meter^3})} \\[0.5ex]
\hline 
\centering Teflon release film  & 0.75 & 0.75 & 0.75 & 0.275 & 0.275 & 0.275 & 0.45 & 0.45 & 0.45 & 2200\\[1ex]
\centering Crushed core  & 0.0214 & 0.0214 & 0.136 & 0.0507 & 0.071 & 0.152 & 0.318 & \SI{3.18e-5}{} & \SI{3.18e-5}{} & 92\\[1ex]
\centering High density core  & 0.238 & 0.238 & 2.0 & 0.368 & 0.515 & 1.105 & 0.318 & \SI{3.18e-5}{} & \SI{3.18e-5}{} & 184\\[1ex]
\hline
\end{tabular}
}
\end{table*}

\subsection{Feature extraction}

Features computed from time-domain signals acquired using the embedded systems are best suited for implementing ML models for damage inference at the edge. For the damage classification model presented in this work, we utilized $10$ features computed from the time-domain signals, summarized in Table \ref{tab:time_features}. Widely used statistical features such as mean absolute deviation (MAD), variance (VAR), standard deviation ($\sigma$) are computed using the monitoring signal alone 
On the other hand, features such as root mean square energy (RMS), root mean square deviation (RMSD) and peak to peak amplitude deviation (PPAD) of signals \cite{SAWANT2021106439}, 
signal energy ratio (SER) \cite{Torkamani_2014} etc. capture the deviation of monitoring signal from baseline signal, and therefore the impact of damage on amplitude and phase of the GW signals. Along with features easily computed from time domain signals, we also computed spectral difference deviation (SDD) using Fourier domain representation. SDD quantifies the difference in the Fourier spectra of baseline ($S_b( f )$) and monitoring ($S_m( f )$) signals, normalized to their respective
energies \cite{ren2019gaussian}.

\subsection{Filtering-based feature selection}

\begin{figure}[!htbp]
    \centering
    \includegraphics[width=0.95\linewidth]{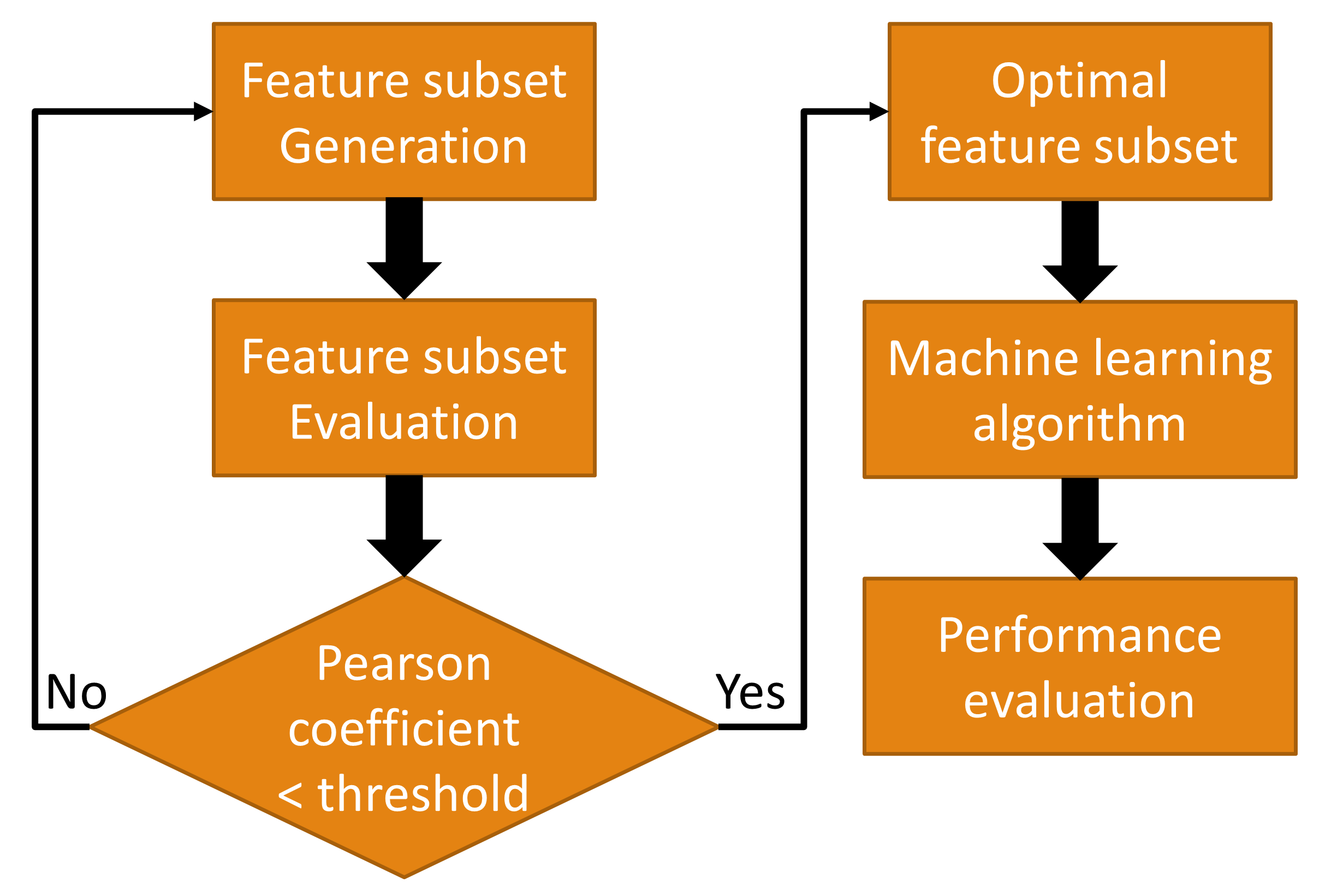}
    \caption{Filtering-based feature selection approach}
    \label{fig:FE_filter}
\end{figure}

Filtering-based approaches are used as commonly used as preprocessing step in ML-pipelines. These methods select features from the original feature set irrespective of the use of which ML algorithm is finally used. In terms of computation, these feature selection methods are fast and computationally inexpensive ways to remove duplicated, correlated and redundant features. Statistical measures used for feature selection in filtering-based methods include Pearson's correlation coefficient for continuous and Chi-square test for discrete features. As shown in Figure\ref{fig:FE_filter}, in each iteration, we consider two out of $10$ features and compute Pearson's correlation coefficient as a feature evaluation metric. We set the threshold on Pearson's correlation coefficient values for selecting the features. Considering subset of features and applying threshold criteria if N features are correlated among themselves then N-1 features are eliminated.  
The optimal feature subset thus obtained is used to train and evaluate the performance of ML models.

\subsection{Damage classification}

Damage classification is modeled in this study as a supervised task using optimal feature subset as input. In a classification problem, the target or dependent variable, $y$, can take only discrete values for a given set of features or independent variables, $X$. We evaluated the performance of five different classifiers : 
\begin{itemize}
    \item Logistic regression:  Logistic regression is typically used for predicting binary targets, it can be extended to multi-class classification problem using heuristic method called "one versus rest", wherein, for $N-$class problem, $N$ different binary classifiers are trained to predict one class as positive and other classes as negative. The coefficients in logistic regression are estimated using maximum-likelihood estimation. Logistic regression is easier to implement, efficient to train and can interpret model coefficients as indicators of feature importance. The major limitation of Logistic regression is the assumption of linearity between the targets and the features. It is also sensitive to class imbalance problem and scaling of features and requires careful tuning.
    
    \item Support vector machine (SVM): For linearly separable classes, the SVM algorithm learns the decision boundary as a hyperplane. Samples closest to the hyperplane from both classes are called support vectors. Its learning objective is to maximize the margin, i.e., the distance between the hyperplane and the support vectors. SVM can be extended to a multi-class problem using one versus one strategy. Here, for N-class problem, a total of $N \times (N-1)/2$ binary classifiers are required, one for every possible pair of classes. If the data is not linearly separable, kernel SVM technique projects data in lower dimensions to higher dimensions in such a way that data points belonging to different classes are allocated to different dimensions and become linearly separable. Training SVMs and tuning its hyperparameters is expensive and they are difficult to interpret compared to simple logistic regression models. 
    
    \item Naive Bayes: It is a probabilistic classifier inspired by the Bayes theorem under a simple assumption that features are conditionally independent. Despite this oversimplified assumption, naive Bayes classifiers work very well in many complex real-world problems. A big advantage of naive Bayes classifiers is that they only require a relatively small number of training data samples to perform classification efficiently, compared to other algorithms, such as logistic regression, decision trees, and SVMs. 
    For this study we use Gaussian naive Bayes wherein values associated with each feature are assumed to be distributed according to a Gaussian distribution. The drawback of calculating model parameters of Gaussian distribution is that we cannot interpret feature importance in this case.
    
    \item Decision tree: It utilizes an if-then rule set which is mutually exclusive and exhaustive for classification in the form of a tree structure. Decision tree uses purity-based loss functions, such as Gini impurity and entropy, to evaluate the split based on the class purity of the resulting node \cite{myles2004introduction}. Features in the top of the tree have more impact towards the classification. Decision tree models have better interpretability. However, decision trees are prone to overfitting. A single decision tree by itself usually has subpar accuracy but the tree structure makes it easy to control the bias-variance trade-off.
    
    \item Random forest: A random forest consists of a large number of individual decision trees that operate as an ensemble. Using bootstrap aggregation or bagging technique, random bootstrapped samples are drawn from training set \cite{breiman2001random}. Further, for each decision tree random subsets of features are used, which often result in better predictive performance than a single decision tree due to better variance-bias trade-offs \cite{breiman2001random}. 

\end{itemize}

\begin{figure}[!t]
    \centering
    \includegraphics[width=0.9\linewidth]{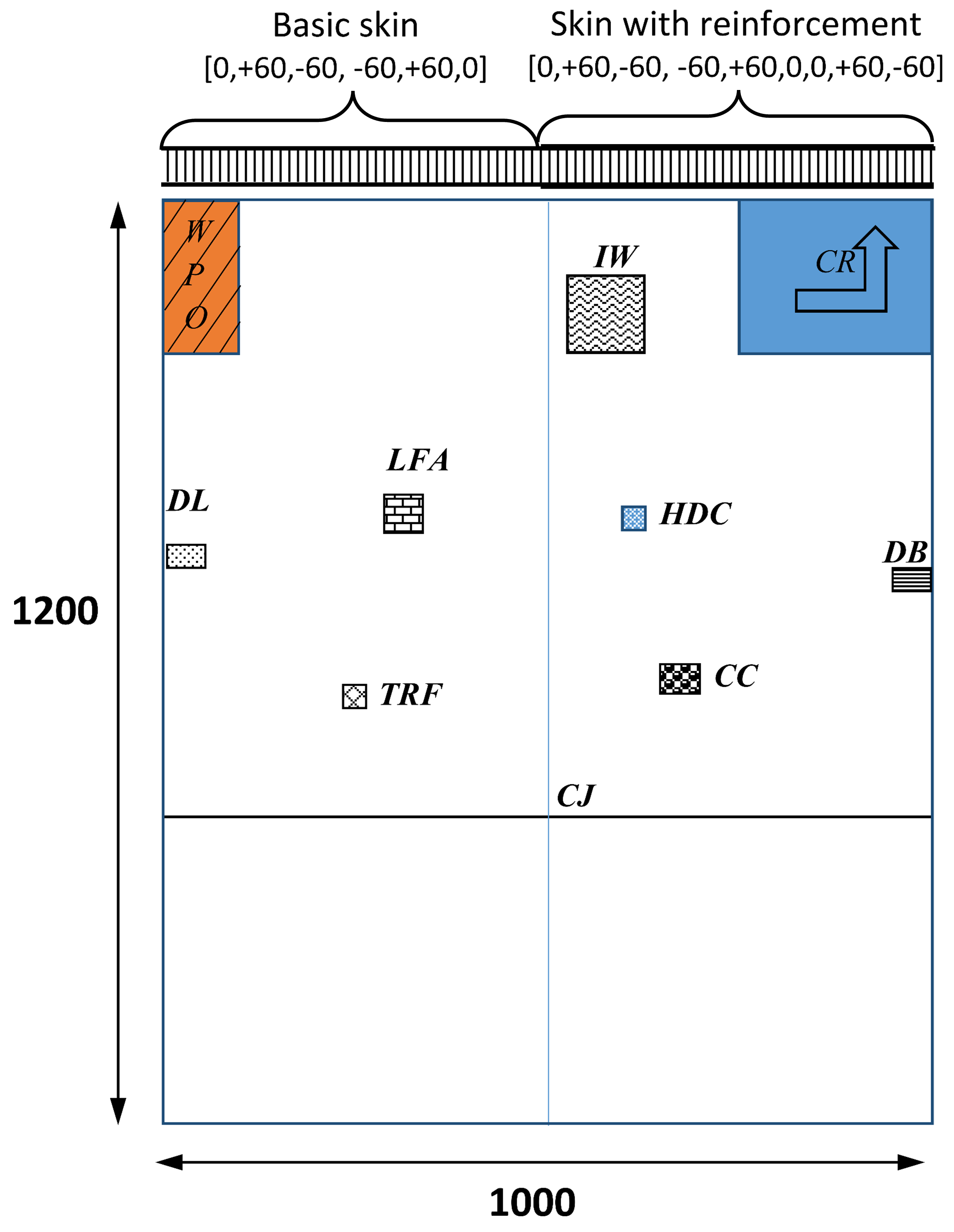}
    \caption{HCSS panel with lateral dimensions \qtyproduct{1000x1200}{mm} containing various types of damages. Out of these, four types of damages, denoted as TRF, LFA, HDC, CC of size \qtyproduct{30x30}{mm} were studied in this work.}
    \label{fig:ISROpanel}
\end{figure}

\begin{figure}[!t]
    \centering
    \includegraphics[width=0.45\linewidth]{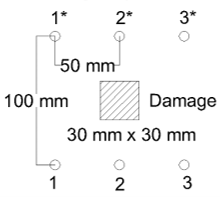}
    \caption{Experimental layout of three Nano-30 sensor paths for the study of four types of damages using guided wave propagation in HCSS panel}
    \label{fig:Nano30_layout_expt}
\end{figure}

\section{Numerical simulation of different types of damages in HCSS}\label{sec:FEM_COMSOL}

\begin{figure}[!t]
     \centering
     \includegraphics[width=0.9\linewidth]{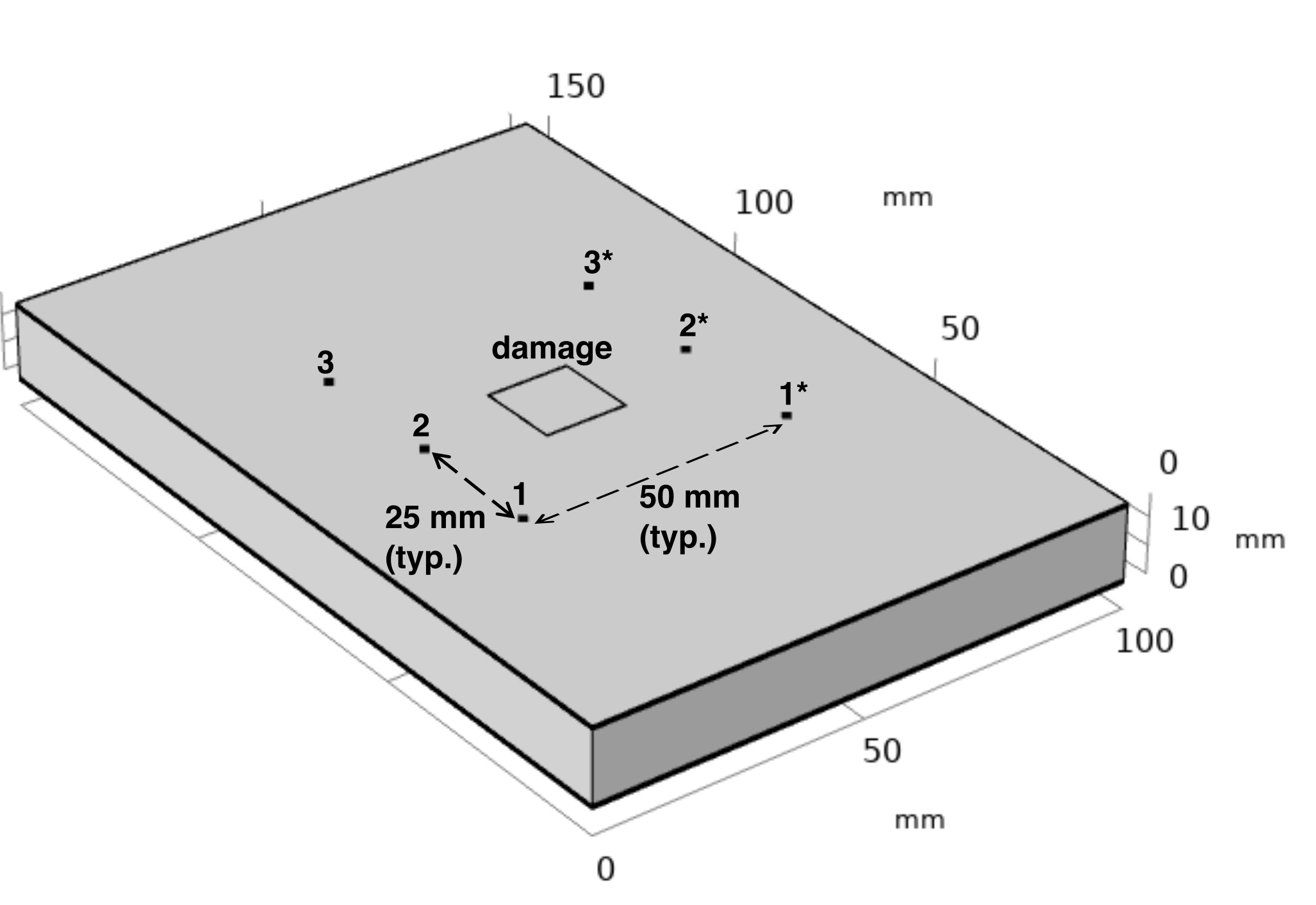}
     \caption{Three-dimensional half-scale FE model of size \qtyproduct{100x150}{mm} with a \qtyproduct{15x15}{mm} damage which falls directly in the wave propagation path P2-2*}
     \label{fig:3DFEM}
 \end{figure}

To study the impact of different types of damages on GW propagation, we carried out numerical simulation on the HCSS panel as shown in Figure \ref{fig:ISROpanel}, using commercially available COMSOL Multiphysics 5.5 software using \textit{Structural Mechanics} module and analyzed it using \textit{Time Implicit} study.
The HCSS panel consists of two halves. One half is of \SI{14.97}{mm} thickness comprising of \SI{1.125}{mm} thick $9-$layer CFRP lamina skin on top and bottom surfaces, with [0/+60/-60/-60/+60/0/0/+60/-60] layup bonded using \SI{0.01}{mm} thick adhesive layer (HEXCEL-Redux 319L) to a \SI{12.7}{mm} thick honeycomb core (HEXCEL-Al $5056$ hexagonal aluminum Nomex core of cell size \SI{0.25}{inch}). The other half is of \SI{14.22}{mm} thickness comprising of \SI{0.75}{mm} thick $6-$layer CFRP lamina with [0/+60/-60/-60/+60/0] layup bonded to the core. Table~\ref{tbl:scsprop} provides the material properties of the sandwich composite structure. 
A half-scale 3D model was used to study the effect of four damages on a \qtyproduct{100x150}{mm} HCSS panel with a \qtyproduct{15x15}{mm} damage as shown in Figure \ref{fig:3DFEM}. GW paths 1-1* and 3-3* (Figure \ref{fig:Nano30_layout_expt}) were considered to be baseline as the damage lies directly along path 2-2*. Out-of-plane load (z-direction) was given as a point load to simulate the actuation produced by the Nano-30 detachable miniature transducers, thus predominantly generating $A0$ mode in the panel, and output out-of-plane displacement is measured at a point \SI{50}{mm} apart. Accuracy of numerical simulation depends on the mesh size and time-step increment in \textit{Time Implicit} study \cite{LECKEY2018187}. As with previous studies on COMSOL based guided wave numerical simulation, the largest mesh element size was restricted to $1/6^{th}$ of the wavelength of the propagating mode \cite{lei2019multiphysics}. The time step was chosen to be \SI{0.1}{\micro s} $(>1/60 f)$. For excitation frequency $f=$ \SI{100}{\kilo Hz}, the wavelength of $A0$ mode in the HCSS panel was approximately \SI{12}{mm}. Thus, the largest mesh size in the plate was set to \SI{2}{mm}. The time-domain profile of the out-of-plane point load was specified as a $5-$cycle sine pulse shaped by a Hanning window, expressed as:
\begin{equation}
 V(t)=0.5\left[1-\cos\left(\frac{2\pi f t}{5}\right)\right]\sin(2 \pi f t)
 \label{EQ:Hann5}
\end{equation}

 \begin{figure}[!t]
     \centering
     \includegraphics[width=0.9\linewidth]{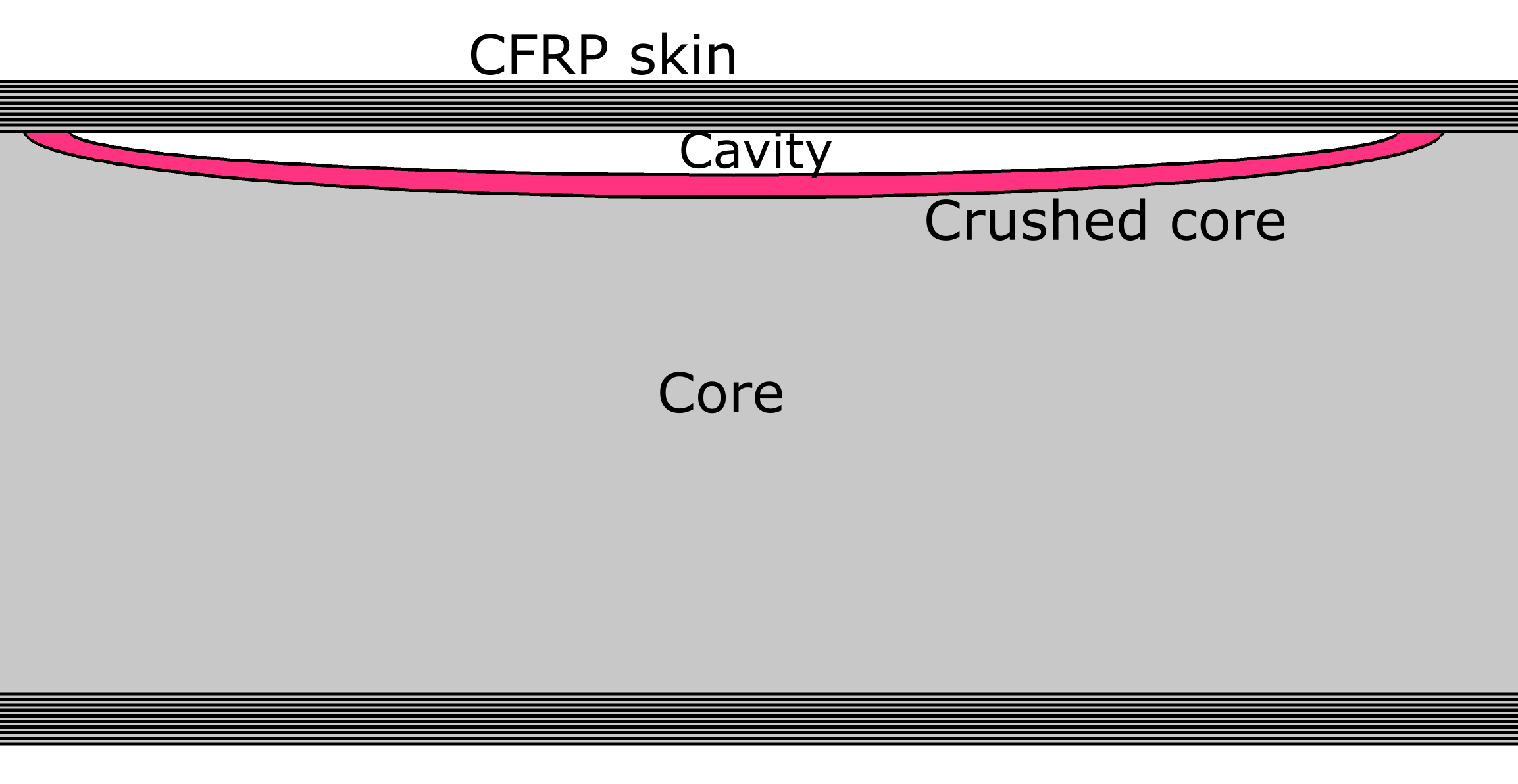}
     \caption{Two-dimensional Core crush (CC) FE model with an ellipsoidal cavity of size \qtyproduct{30x1}{mm} with a crushed core}
     \label{fig:CCFEModel}
 \end{figure}

\begin{figure}[!t]
     \centering
     \includegraphics[width=0.9\linewidth]{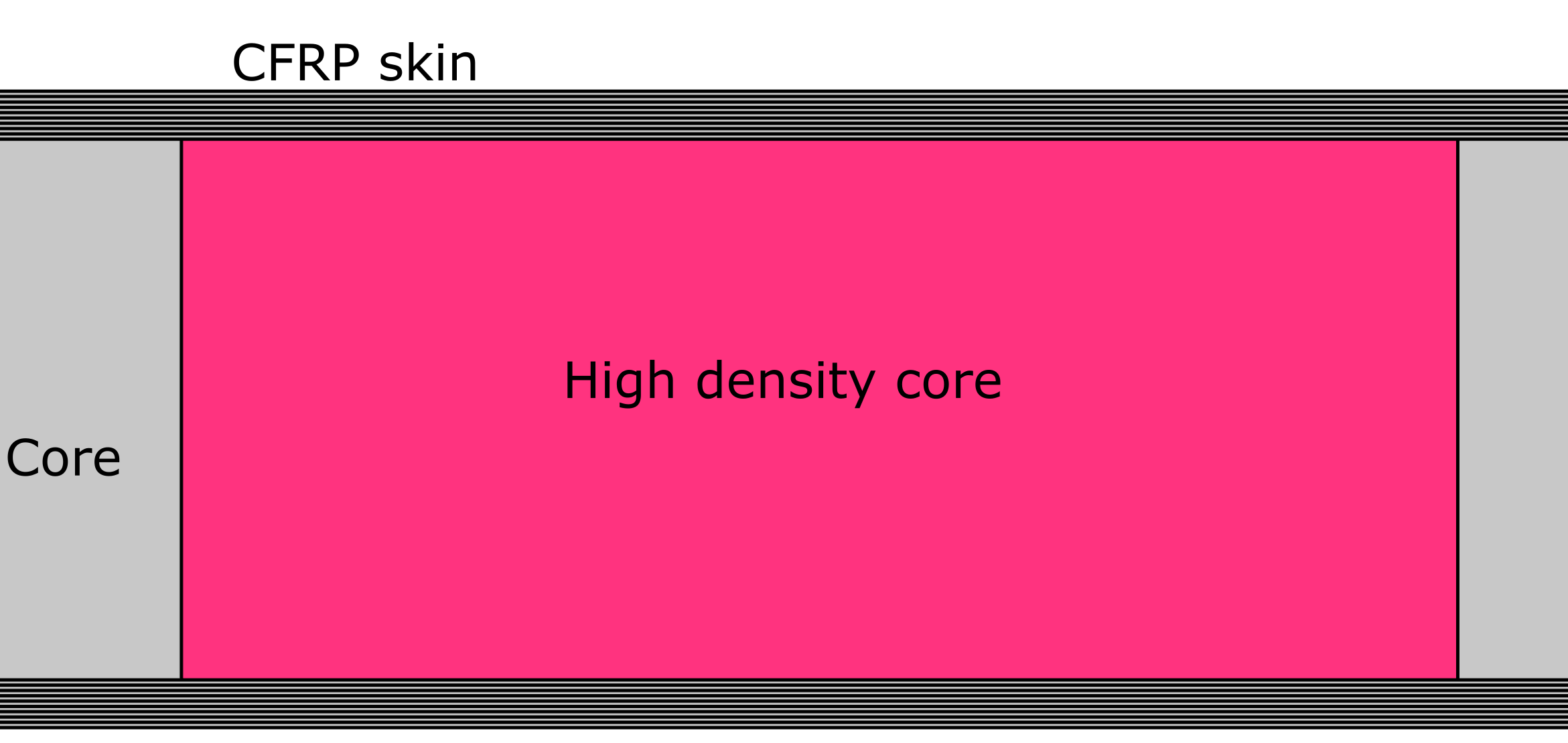}
     \caption{Two-dimensional high density core (HDC) FE model with a \qtyproduct{30x12.7}{mm} HD core zone}
     \label{fig:HDCFEModel}
 \end{figure}
  
\begin{figure*}[!t]
    \centering
    \includegraphics[width=0.85\linewidth]{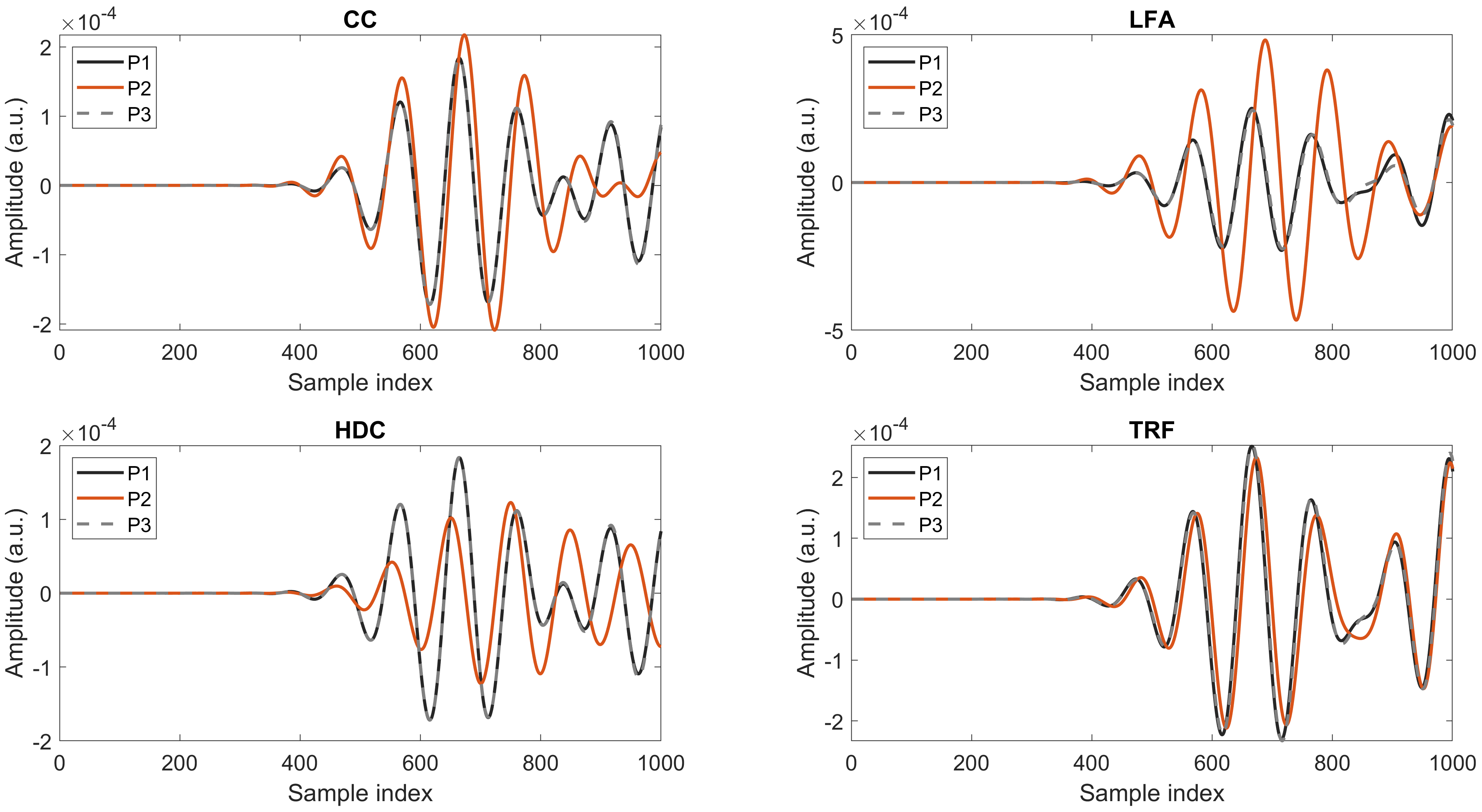}
    \caption{GW signals for four types of damages corresponding to three paths as per layout of the half-scale model shown in Figure~\ref{fig:3DFEM}} 
    \label{fig:GWFEM3D}
\end{figure*}

\begin{figure}[!t]
     \centering
     \includegraphics[width=\linewidth]{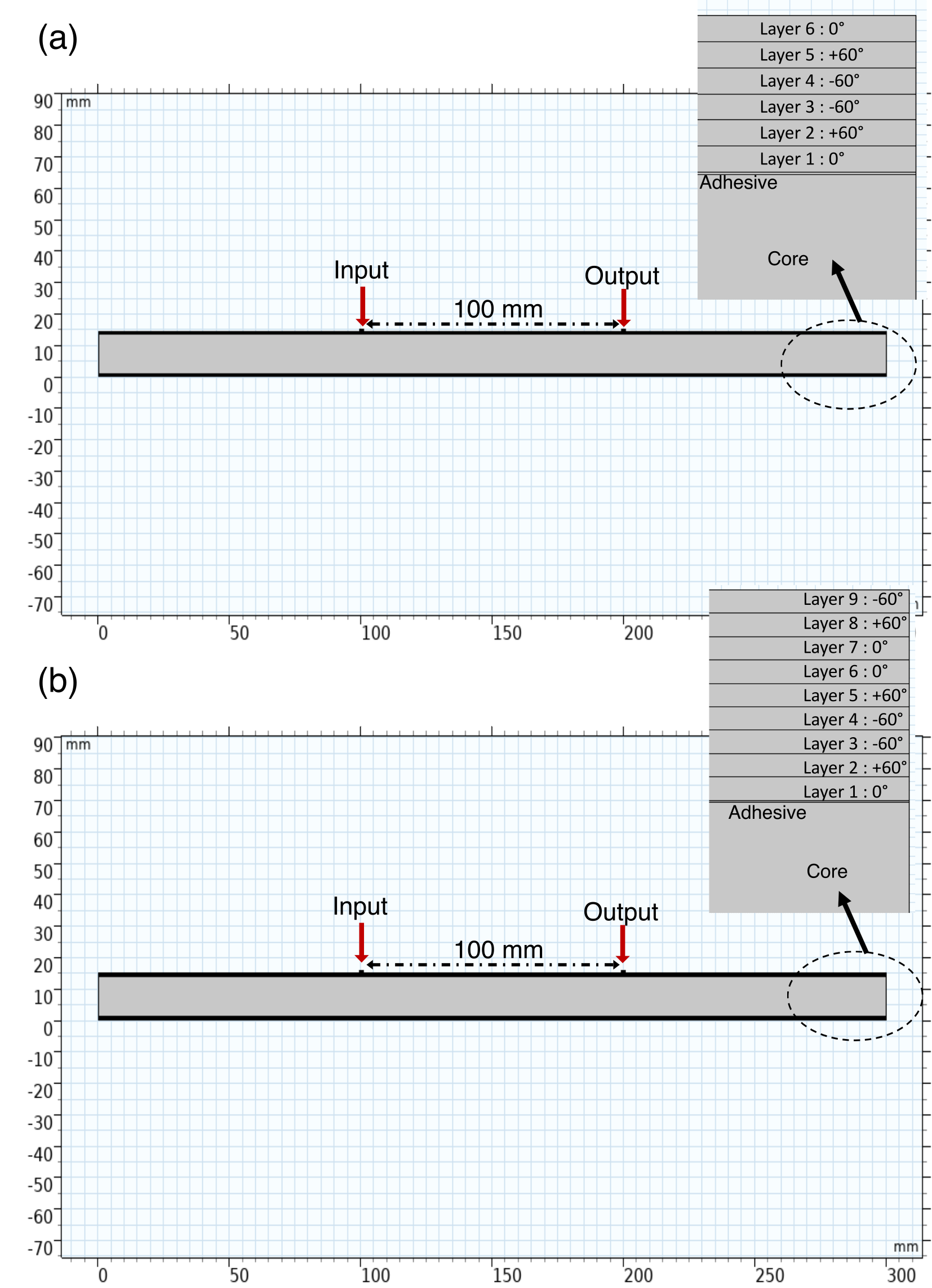}
     \caption{Two-dimensional FE model of (a) size \qtyproduct{300x14.22}{mm} comprising \SI{0.75}{mm} thick $6-$layer CFRP lamina skin on top and bottom surfaces with [0/+60/-60/-60/+60/0] layup bonded using \SI{0.01}{mm} thick adhesive layer (HEXCEL-Redux 319L) to a \SI{12.7}{mm} thick honeycomb core \& (b)\qtyproduct{300x14.97}{mm} comprising \SI{1.25}{mm} thick $9-$layer CFRP lamina skin on top and bottom surfaces, with [0/+60/-60/-60/+60/0/+60/-60] layup}
     \label{fig:2DFEM}
 \end{figure}

\begin{figure*}[!t]
    \centering
    \includegraphics[width=0.8\linewidth]{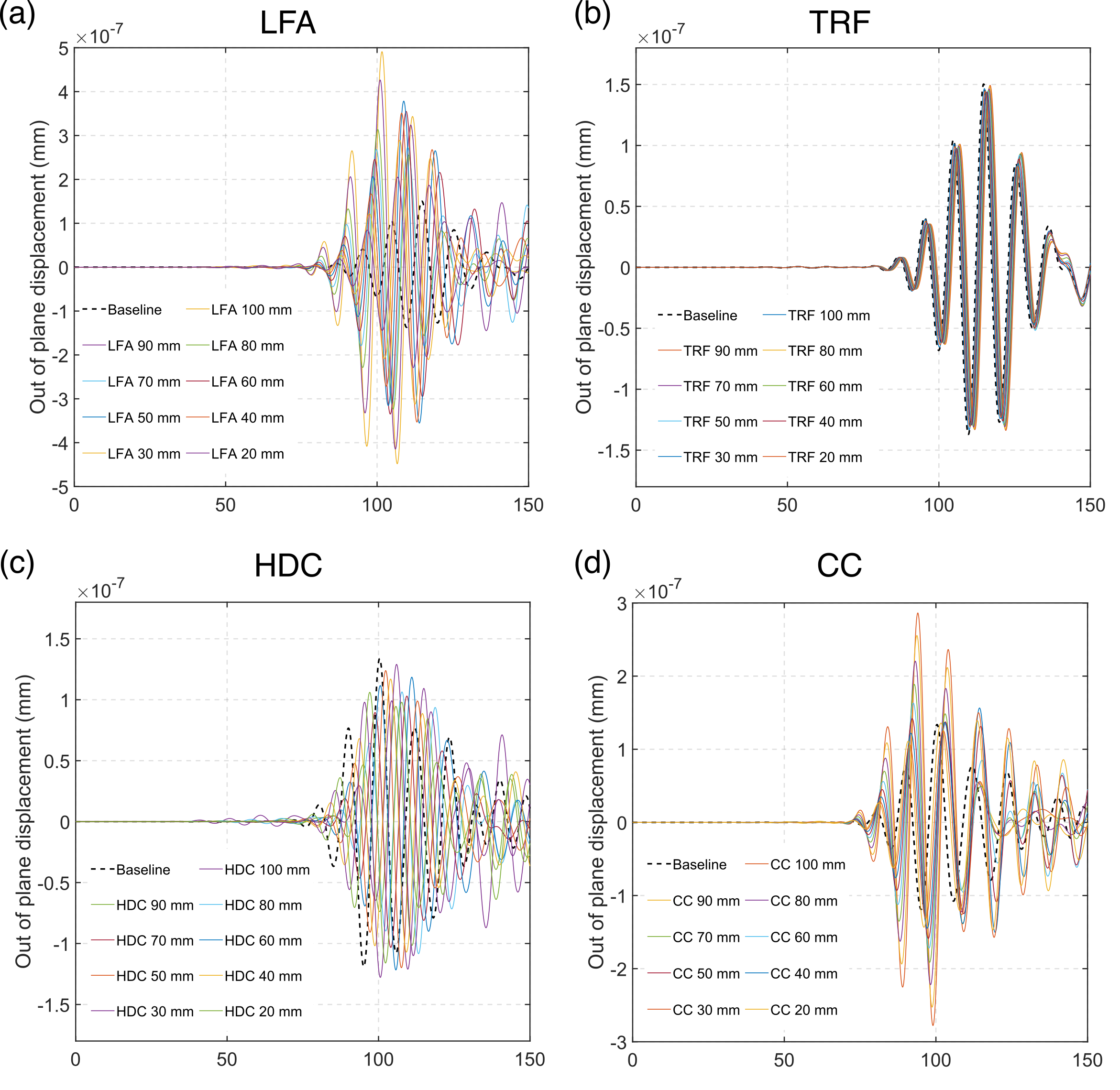}
    \caption{GW signal comparing baseline versus damaged HCSS panel with increasing damage parameters for (a) LFA (b) TRF (c) HDC \& (d) CC }
    \label{fig:GWFEM}
\end{figure*}
 
 To study the effect of damages present in the HCSS panel, signals for baseline and the path containing the damage signal were compared for four types of damages introduced in the model. LFA of length "L" was modeled by removing the adhesive layer between the CFRP skin and the honeycomb core, whereas TRF was modeled as a \SI{0.01}{mm} thick Teflon film in between the individual lamina. The material properties of Teflon considered in the simulations are tabulated in Table \ref{tbl:damageprop}. In the other half of the panel, CC was modeled by creating a cavity with a crushed core region beneath the cavity at the skin-core interface. The cavity and crushed core were modeled as a half-ellipsoidal surface as shown in Figure \ref{fig:CCFEModel}. Previously reported experimental investigations \cite{10.1115/1.4055549} were analyzed to deduce the crushed core material properties and are tabulated in Table \ref{tbl:damageprop}. HDC zone of thickness \SI{12.7}{mm} present in HCSS core as shown in Figure \ref{fig:HDCFEModel} was modeled in 2D FEM simulations with the material properties as listed in Table \ref{tbl:damageprop}.

Figure \ref{fig:GWFEM3D} shows that the time-series data for the path 2-2* containing the damage has a significant effect on the amplitude with respect to the baseline paths 1-1* and 3-3*. It is evident from the 3D simulations that CC and LFA damage signals shows an increase in the peak amplitude when compared to the baseline while HDC and TRF shows a decrease in amplitude.

\hil{\subsection{Parametric study} \label{sec:parstudy}
 Generating experimental data poses a significant challenge due to the high costs and impracticality of creating different types of damages. To evaluate the performance of the proposed algorithm for different damage sizes, we also conducted extensive parametric study using two separate 2-D numerical simulations, which were performed as follows:}
\begin{itemize}
    \item A \qtyproduct{300x14.22}{mm} thick HCSS panel to study the effect of LFA and TRF as shown in Figure\ref{fig:2DFEM}(a).
    \item A \qtyproduct{300x14.97}{mm} thick HCSS panel to study the effect of CC and HDC as shown in Figure\ref{fig:2DFEM}(b).
\end{itemize}

The studies carried out by varying parameters and GW signal data from numerical simulation of baseline and damaged path were compared for the four type of damages. Figure \ref{fig:GWFEM} shows that the time-series data for a damaged path has a significant effect on the amplitude with respect to baseline signal. It is evident from the various case studies presented that CC and LFA damage signals show an increase in the peak amplitude when compared to the baseline, while HDC and TRF show a decrease in amplitude. \hil{Due to the large computational cost associated with 3D modeling, we used 2D COMSOL models for the parametric study, a thorough analysis using 3D models is required to investigate effect of damage size on amplitude and phase of the baseline signal.}

\begin{figure}[!t]
    \centering
    \includegraphics[width=0.7\linewidth]{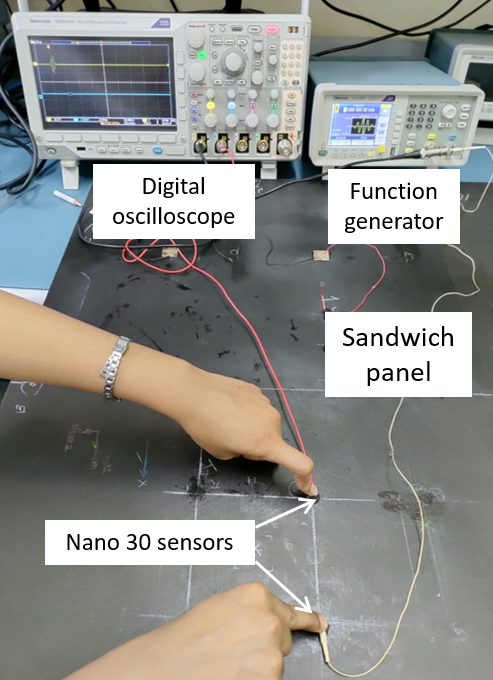}
    \caption{Experimental setup showing HCSS panel instrumented with a pair of Nano-30 detachable transducers used as actuator (transmitter) and receiver, function generator for generation of actuation signal, and oscilloscope for recording the receiver response.}
    \label{fig:exp_setup}
\end{figure}

\begin{figure*}[!t]
    \centering
    \includegraphics[width=0.85\linewidth]{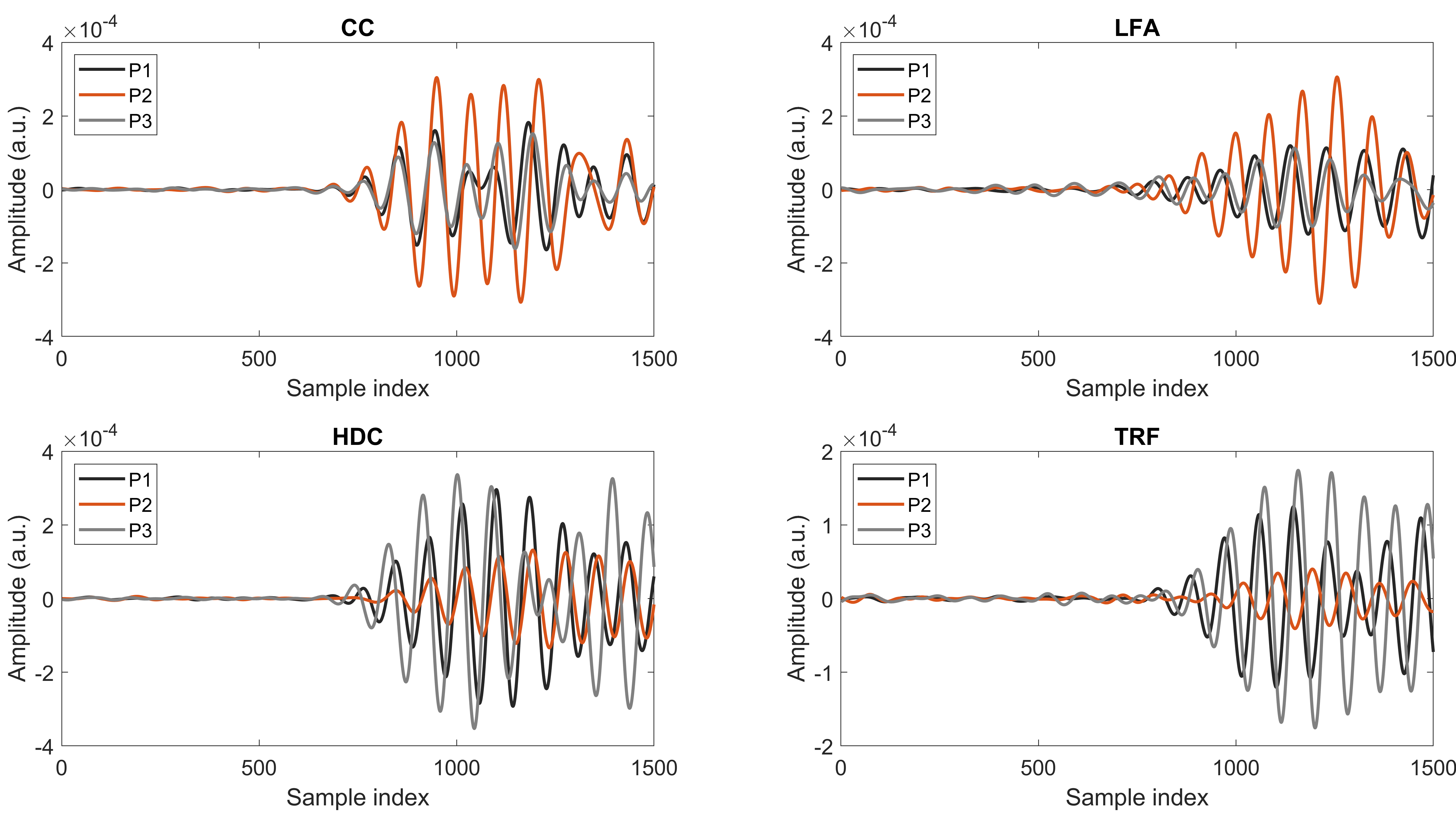}
    \caption{GW signals for four types of damages corresponding to three paths as per layout shown in Figure~\ref{fig:Nano30_layout_expt}. For experimental study, we recorded \SI{20}{} such time-series data trials per path for each of four damages.}
    \label{fig:Allsig}
\end{figure*}

\begin{figure*}[!t]
    \centering
    \includegraphics[width=0.8\linewidth]{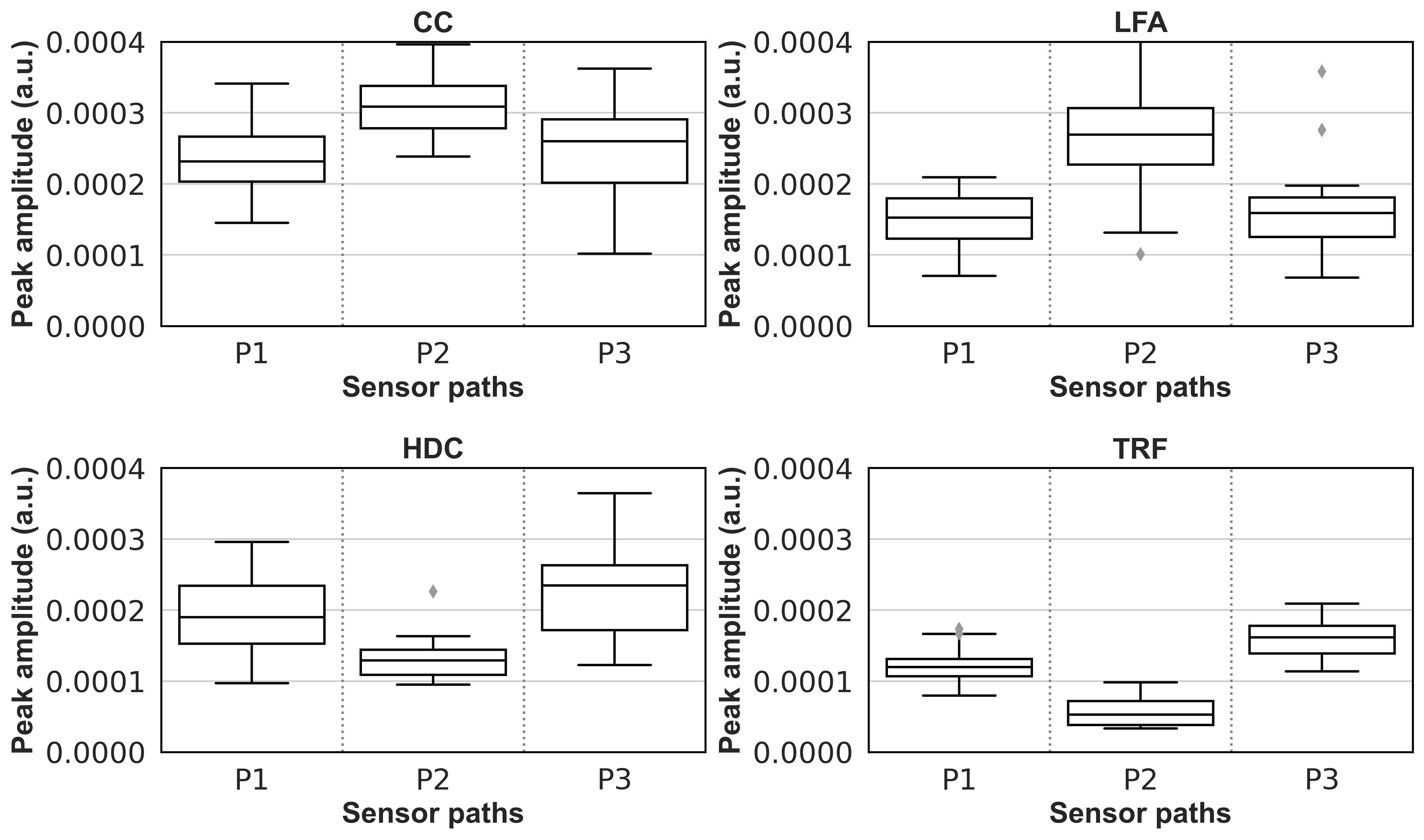}
    \caption{Trends in peak amplitude values of four damages corresponding to twenty trials for each of the three paths recorded using a pair of detachable Nano-30 transducers}
    \label{fig:PeakTrends}
\end{figure*}

\section{Results and discussion}

\subsection{Experimental setup}

\begin{figure*}[!t]
    \centering
    \includegraphics[width=0.9\linewidth]{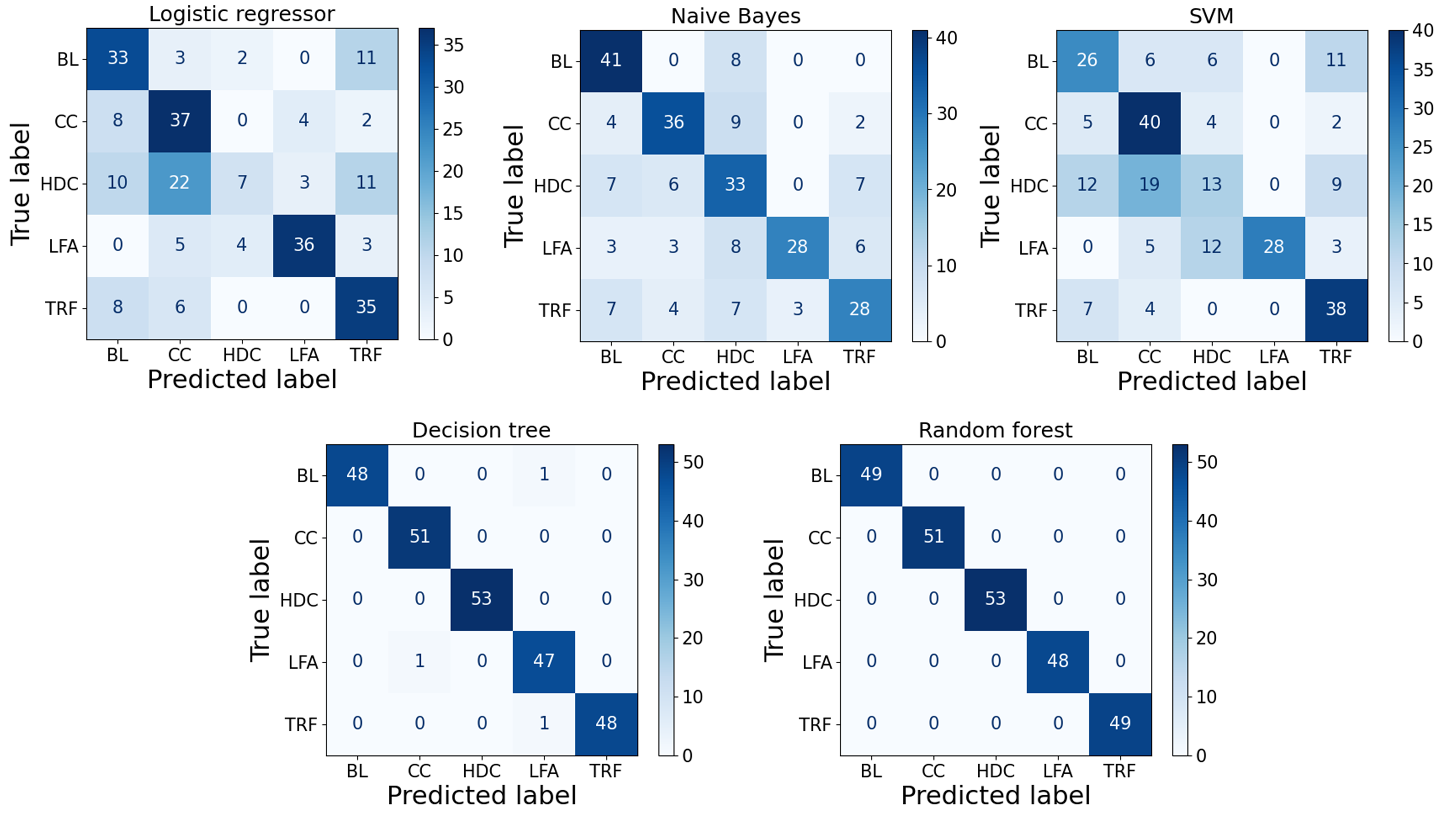}
    \caption{Confusion matrices for the five classifiers on held out experimental data}
    \label{fig:Conf_mat}
\end{figure*}

\begin{figure*}[!t]
    \centering
    \includegraphics[width=0.9\linewidth]{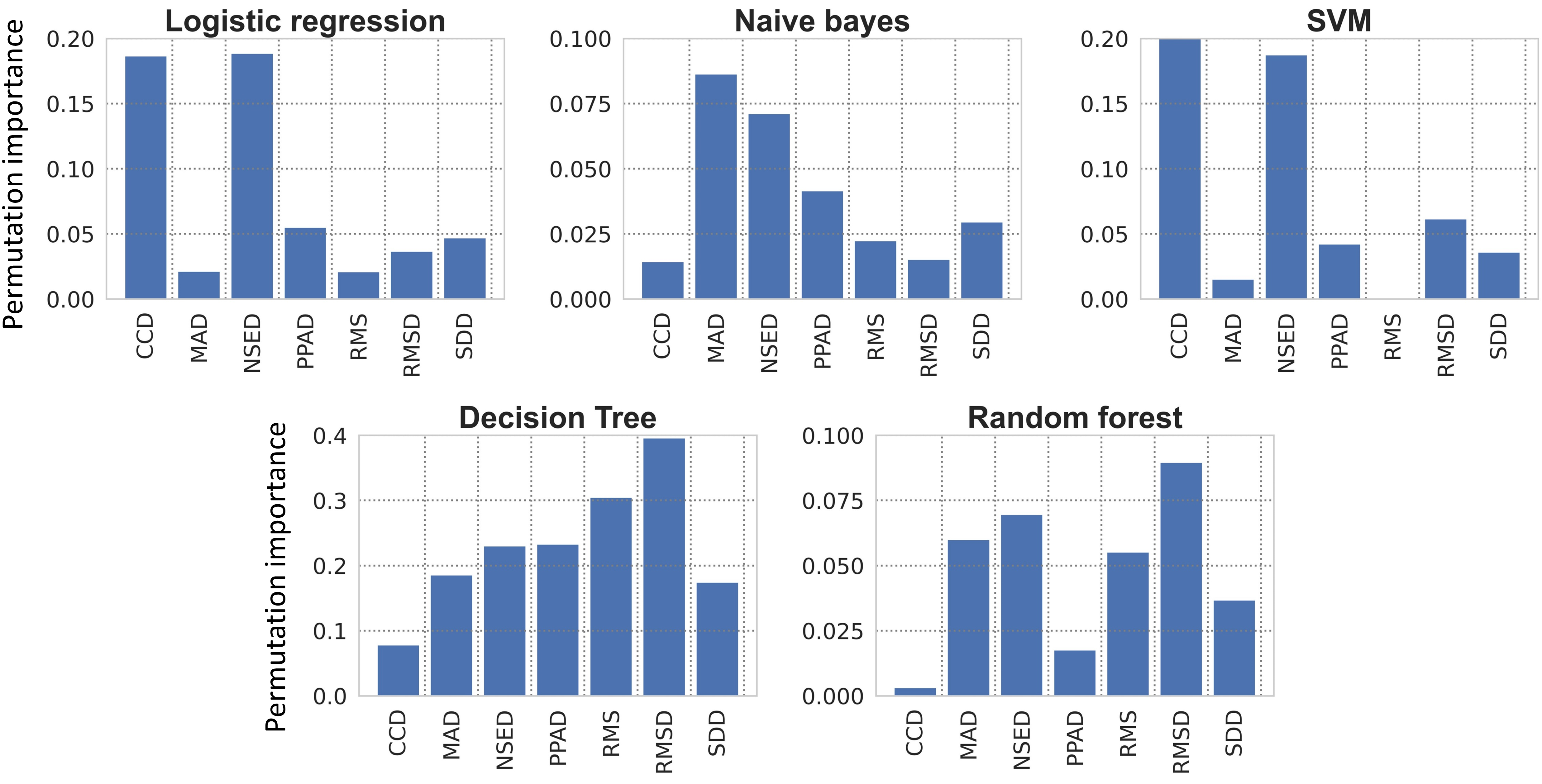}
    \caption{Feature importance using permutation technique for all classifiers for experimental data}
    \label{fig:FE_importance}
\end{figure*}

To study the change in baseline signal due to different types of damages, we analyzed a HCSS panel as shown in Figure~\ref{fig:ISROpanel}. To perform non-destructive ultrasonic GW testing of the HCSS panel with various damages, the experimental setup shown in Figure~\ref{fig:exp_setup} was used. We used a pair of Nano-30 transducers (separated by \SI{100}{mm} distance) for GW transduction using commercially available ultrasound gel as couplant. Compared to conventionally used piezoelectric wafer transducers, these contact-type transducers do not need to be permanently attached to the HCSS surface with an adhesive, and are thus suitable for non-destructive testing. These transducers generate predominantly anti-symmetric modes, and the interaction of the fundamental anti-symmetric $(A0)$ mode with various damages depends on the type of defect. A function generator (Tektronix AFG1022) was used to generate $5-$cycle sine pulse shaped with Hanning window as actuation signal (equation \ref{EQ:Hann5}). The signal frequency was \SI{100}{\kilo Hz} and peak-to-peak voltage amplitude was \SI{20}{V}. A digital storage oscilloscope (DSO, Tektronix MDO3024) was used to record the GW signal acquired by the receiver Nano-30 at a sampling rate of \SI{100}{MHz}. The amplitude of the recorded signal is affected by various parameters, such as acoustic coupling, mechanical impedance, position of the transducers, and pressure applied on the transducers. The Nano-30 transducers were attached manually on the panel as shown in Figure~\ref{fig:exp_setup}, that introduced variation in the amplitude. Therefore, \SI{20}{} measurements were recorded for each of the three paths for each damage with the transducers removed and attached again for each measurement. The positions of the sensor paths relative to the defect are shown in Figure \ref{fig:Nano30_layout_expt}. Note that the region of the panel containing the damage falls directly along the wave propagation paths P2-2*. GW response for path P1-1* and 3-3* were considered as baseline signal, since it is sufficiently separated from the damages. 

Recorded time-series data contains an offset. Discrete wavelet transform (DWT)-based approach was used to preprocess the GW signals prior to computation of features. DWT yields step-wise decomposition of a signal in different frequency bands centered at a specified frequency scale. For this purpose, Daubechies mother wavelet of order $40$ was used to obtain a $7-$level wavelet decomposition of each signal, and the $6^{th}$ coefficient was used as the filtered version of the signal. Detailed discussion on the DWT-based denoising method can be found in work by Rizzo et al. \cite{rizzo2005ultrasonic}. 

Representative time-series data for one of the \SI{20}{} trials thus obtained for the three sensor paths for the four damage types is illustrated in Figure~\ref{fig:Allsig}. \hil{Experimental data was observed to be in accordance with simulation data (Figure~\ref{fig:3DFEM}), which confirms that the experimental data was a good representation of the underlying physical model.} We observe that time-series for sensor path P2-2*, containing the damage, shows significant change in the amplitude with respect to sensor paths P1-1*, and P3-3*, which are relatively away from the damage hence used as baseline data. It is important to note that two damages, CC and LFA result in decrease in amplitude with respect to baseline, whereas the remaining two damages HDC and TRF, show decrease in amplitude with respect to baseline.

Trends in peaks of the \SI{20}{} trials recorded using detachable Nano-30 transducers in shown in Figure~\ref{fig:PeakTrends}. These trends are consistent with the variation of the signal containing the damage (P2-2*) with respect to baseline data (P1-1* and P3-3*). The challenge here is to distinguish between damages that cause similar change in the baseline signal.

To incorporate uncertainties due to environmental and operational conditions, we added white noise to each time-series $s(t)$ :

\begin{equation}
s_{n}(t) = s(t)+ \beta_n \max(s(t))\left[w(t)\right]
\label{EQ:Noise}
\end{equation}
where $s_{n}(t)$ denotes the signal with added noise, $\max(s(t))$ is the peak amplitude of signal $s(t)$, $\beta_n$ denotes the noise scaling factor, and $w(n)$ is white noise generated using normal distribution with zero mean and standard deviation of $1$. We studied the performance of the damage classification method for signal to noise ratio (SNR) value of \SI{25}{dB} using $\beta_n=0.01$. 
Each experimental GW time-series obtained after wavelet-based preprocessing was used to generate noise-augmented data by adding white noise as described in equation \eqref{EQ:Noise}. For feature computation and subsequent damage assessment, 10 noisy copies of each time-series were generated randomly. After noise augmentation, there are total 200 time-series data for each of four damage types recorded for sensor path P2-2*. For baseline data, we have data corresponding to P1-1* and P3-3* for all four damages, i.e. total 80 time-series. We performed the study by selecting 20 baseline time-series samples from different damage types and creating 10 noisy copies of each time-series,  it doesn't affect the classification accuracy. Here we have reported the study using 10 baseline samples from CC damage and 10 samples from TRF damage, total 20 samples are then noise augmented to get 200 samples with SNR of 25 dB. 


\subsection{Feature engineering}
As listed in Table\ref{tab:time_features}, we computed total 10 features for each noisy time-series. For computation few of features such as RMSD, PPAD, SER etc., we used time-series data corresponding to P3-3* as baseline $f_b(t)$. At the end of feature extraction step, we obtained feature set with dimension $1000\times10$. 
Next, out of the 10 features, the most relevant features were selected using filtering approach with Pearson's correlation coefficient as selection criteria. Correlation is a measure of the linear relationship of two features. The logic behind using correlation for feature selection criteria is that the most relevant features are highly uncorrelated among themselves. If two features are correlated, we can predict one from the other. Therefore, an ML model only needs one of them, as the second one does not add additional information. We set an absolute value of 0.95 as the threshold. Using thresholding, if in the subset of N features are correlated among themselves then N-1 features are eliminated.  


\subsection{Damage classification}

For the damage classification problem, we used the optimal feature subset of size $1000\times 7$ obtained in the previous step as our input and targets were generated in one hot encoding scheme corresponding to five classes : Baseline, CC, LFA, HDC and TRF. The dataset was randomly split in the proportion $75\%:25\%$ for training and testing, respectively. We evaluated performance of five different ML models for five class classification. We used scikit-learn library for implementing the machine learning in python. \hil{The codes are open sourced on \href{https://github.com/shrutisawant099/Damage-Classification-using-feature-engineering.git}{GitHub}. We repeated the experiment with 10 different random splits i.e. randomly selecting $75\%:25\%$ for train:test dataset in each trials. Accuracy as average of 10 trials for each of the five models is listed in Table\ref{tab:Acc}.} Logistic regressor and SVM have lower accuracy as these models work better for linearly separable data. Gaussian naive Bayes performs comparatively better for non-linear data due to its probabilistic approach. Random forest performs better than decision tree by considering multiple decision trees at a time. Confusion matrices are shown in Figure \ref{fig:Conf_mat}.

To understand the effect on each feature on the performance of the model we used permutation importance metric. It shuffles the data corresponding to a feature and computes the drop in accuracy (called as permutation importance) using test data with a trained model. If a particular feature results in a significant drop in the accuracy then that feature has more importance in the prediction of the model. This metric is especially beneficial in case of Gaussian naive Bayes model where we get model parameters for Gaussian distribution and not weights of each feature. 
Permutation importance has also been used for highly complex or non-parametric models such as support vector machines. It addresses the issue where random forest models are biased towards categorical variables with a large number of categories \cite{altmann2010permutation}.
Figure \ref{fig:FE_importance} shows permutation importance of five classifiers for the seven selected features. Here, RMSD plays crucial part in predicting test data with higher accuracy. RMSD computes the change in the amplitude of monitoring signal with respect to baseline. The next highest importance is observed for Fourier domain features such as SDD and NSED. 
Frequency domain decomposition such as Fourier spectrum are best suited to capture the non-stationary nature of GW signals \cite{torbol2014real}.
We also observe that the baseline-free features have relatively lower importance compared the features computed using baseline, which underlines the importance of signal normalization by using the baseline information. Another important observation here is that the decision tree has relatively higher range of importance values. This can be explained based on the tendency of decision tree to have high variance. Random forest using bagging technique fixes this issue and has a better bias-variance trade-off.

\begin{table}[!t]
\centering
\caption{Classification accuracy for various classifiers using the six selected features}
\label{tab:Acc}
\begin{tabular}{|l|r|}
\hline
\textbf{Classifier}    & Accuracy (\%)\\ \hline \hline
Logistic regression  & 59.2 \\
SVM (Linear)  & 58.8 \\
Naive bayes (Gaussian)  & 66.4 \\
Decision trees  & 98.4 \\
Random forest  & 100.0 \\ 
\hline
\end{tabular}
\end{table}


\hil{ \subsubsection{Different damage sizes}
\label{Sec:2dclassify} 

Next, we evaluated feasibility of the proposed feature engineering based damage classification framework for damage classification across different damage sizes. We used simulated data obtained from parametric study reported in section~\ref{sec:parstudy}. From 2D numerical models, we generated one baseline signal and nine timeseries corresponding to nine damage sizes between \SI{20}{mm} to \SI{90}{mm}. For each damage and baseline, we obtained ten noise augmented copies for each damage size, with feature set of size $450\times10$. In this case, we obtained accuracy of 77.89\% indicating robustness of the proposed method for nine damage sizes.  This validates the use of simulation to assess the wider utility of the proposed method. Optimal feature set used here consisted of the same seven features as experimental study.  Lower accuracy value compared to experimental study  can be explained using non-linear relationship between the size of damages and their effect on amplitude and phase of baseline signal as illustrated in Figure~\ref{fig:2DFEM}.}

\subsubsection{Baseline-free feature extraction}
\label{Sec:LiuFeatures}

The baseline-free features reported by Liu et al \cite{liu2021multi} are listed in Table \ref{tab:LiuFeatures}. We computed these 13 features using experimental data for all damages. Noisy data corresponding to P2-2* was used to compute features for four damages and baseline features for P3-3* for two damages CC and TRF. The feature set of size $1000\times13$ was filtered using Pearson's correlation coefficient with threshold of 95\%. Features SF4, SF5, SF6, SF10, SF12 and SF13 were dropped and optimal feature set of dimension $1000\times7$ was used for damage classification. Random forest model gave the highest accuracy of 69.2\% among five classifiers. 
These features are computed using monitoring data alone for pulse echo technique, which generally contains actuation signal along with its reflection. GW data obtained from the network of transmitter and receiver transducers instrumented on composite panels using pitch-catch technique does not consist of an actuation signal. Therefore, for pitch catch GW data, features that compute deviation from baseline work better.





\section{Conclusion and future work}

Damage classification using ML-based approaches reported in GW-SHM literature consider different locations or intensities of the same damage as different classes of ML model. Also, they use deep learning techniques for capturing deviation in the amplitude of the data corresponding to the damaged path from the baseline data, resulting in large model size. The challenge arises when we have limited training data, which is usually the case in a a real world scenario, and we try to distinguish between two or more types of damages in HCSS that cause similar change from the baseline data. In this work, we reported feature engineering-based damage classification using a total 10 statistical features that do not require knowledge of material properties, such as group velocity, time of arrival of wavemode etc. \hil{Therefore, the proposed features and method can be utilized for different materials without knowledge of material properties.} We classified four damages in HCSS, which cause similar changes to baseline data. Optimal feature subset was obtained by dropping four highly correlated features using Pearson's correlation criteria. \hil{If we plug more statistical features in the list, this feature engineering approach will automatically filter out the redundant features without relying on domain knowledge.} Perfect accuracy of damage classification in experimental test data was reported using seven selected features using a random forest classifier. \hil{Also, the robustness of the proposed framework for nine damage sizes is demonstrated using simulation data with an accuracy of 77.89\%.} To the best of our knowledge, this is the first report on damage classification for HCSS using GW-SHM domain. Interpretability of the model using feature importance indicates that features capturing deviation of monitoring signal from baseline signal are more effective compared to baseline-free features. 


In the future we will explore edge-based deployment of these light-weight models inspired by TinyML framework \cite{sanchez2020tinyml}, that could be directly embedded in the FPGA-based embedded system. This will pave the way for designing truly portable, end-to-end smart SHM systems, which can acquire the data and perform damage classification, specifically for efficient in-situ monitoring of HCSS. \hil{In our more recent work, we explored damage classification in the presence of data loss that may occur in wireless sensor networks or IoT-based SHM systems \cite{SAWANT2021106439}. Furthermore, to make the method more robust to real world scenarios, we seek to evaluate the performance of the algorithm for different materials under varying environmental temperatures.}

\begin{table}[ht]
\begin{center}
\centering
\small
\caption{Features computed from GW signals reported by \cite{liu2021multi}, where x is time domain signal and f is frequency domain representation. }

\label{tab:LiuFeatures}
\begin{tabular}{p{1cm}p{5cm}}
\hline
\textbf{Feature}      & \textbf{Expression} \\ \hline  
\\  [0.5ex]

SF1 & $\frac{1}{M}\sum_{i=1}^{i=M}{x(i)}^{3}$ \\ [3ex]

SF2 & $\frac{1}{M}\sum_{i=1}^{i=M}{x(i)}^{4}$ \\ [3ex]

SF3 & max(x) - min(x) \\ [3ex]

SF4 & $\frac{1}{M}\sum_{i=1}^{i=M}{x(i)}^{4}$ / ${(\frac{1}{M}\sum_{i=1}^{i=M}{x(i)}^{4} )^2}$ \\ [3ex]

SF5 & $\sqrt{\frac{1}{M}\sum_{i=1}^{i=M}{x(i)}^{2}}$\\ [3ex]

SF6 & $\sqrt{\frac{1}{M}\sum_{i=1}^{i=M}{(x(i)-\hat{x})}^{2}}$\\ [3ex]

SF7 & max($|x|$) / $\sqrt{\frac{1}{M}\sum_{i=1}^{i=M}{x(i)}^{2}}$\\ [3ex]

SF8 & $\sqrt{\frac{1}{M}\sum_{i=1}^{i=M}{x(i)}^{2}}$ / $\frac{1}{M}\sum_{i=1}^{i=M}\sqrt{|x(i)|}$ \\ [3ex]

SF9 & max($|x|$) / $\frac{1}{M}\sum_{i=1}^{i=M}\sqrt{|x(i)|}$ \\ [3ex]

SF10 & max($|x|$) / ${\frac{1}{M}\sum_{i=1}^{i=M}\sqrt{|x(i)|}}^{2}$ \\ [3ex]

SF11 & $\frac{1}{M}\sum_{i=1}^{i=M}{f(i)}^{2}$ \\ [3ex]

SF12 & $\frac{1}{M}\sum_{i=1}^{i=M}{f(i) - \hat{f}}^{2}$ \\ [3ex]

SF13 & $\frac{1}{M}\sum_{i=1}^{i=M}{f(i)}$ \\ [3ex]

\hline
\end{tabular}
\end{center}
\end{table}


\begin{sm}
The data that support the findings of this study are available
upon reasonable request from the authors. Codes developed for the proposed method are open-sourced on \href{https://github.com/shrutisawant099/Damage-Classification-using-feature-engineering.git}{GitHub}
\end{sm}

\begin{acks}
This work was partially supported by Indian Space Research Organization (ISRO) [grant no. RD/0118-ISROC00-006]. The authors thank Mr.Ramana Raja, PhD student, Department of Civil Engineering, IIT Bombay and  Kajal Shivgan at Wadhwani Electronics Laboratory (WEL), IIT Bombay for assistance with data collection.
\end{acks}








\bibliographystyle{SageH}
\bibliography{refs.bib}

\end{document}